\documentclass[reqno]{amsart}
\usepackage{amsmath, amsthm, amssymb}   % standard math packages
\usepackage{mathrsfs}                   % for script U
\usepackage{mathdots}
\usepackage{color}
\usepackage{doi}
\usepackage{makecell}
\hypersetup{draft}
\usepackage{lineno}
\usepackage{nicefrac,xfrac}
\usepackage{enumitem}                   % for the quotation environment
 \usepackage{filecontents}
 \usepackage[round]{natbib}
\usepackage[T1]{fontenc}

\usepackage{array}
 
 \usepackage{csquotes}
    \setcitestyle{notesep={: }}
  \usepackage[british]{babel}
\usepackage{footmisc}
 
\renewenvironment{quotation}
               {\list{}{\listparindent=0pt
                        \itemindent
                        \listparindent
                        \leftmargin=16pt
                        \rightmargin=16pt
                        \topsep=6pt
                        \itemsep=0pt
                        \parsep=\medskipamount
                       }
                \item\relax}
               {\endlist}

\newcommand{\rN}{\sim_{\textsc{NDV}}}

\begin{document}
\title[Infinite utility: counterparts and ultimate locations]{Infinite utility: counterparts  \\ and ultimate locations}
\author{Adam Jonsson}
\address{Department of Engineering Sciences and Mathematics \\Lule{\aa} University of Technology, Sweden}
\email{adam.jonsson@ltu.se}

%\thanks{Declaration of competing interests: The author declares none.}
%\date{\today}

\begin{abstract}
The locations problem in infinite ethics concerns the relative moral status of different categories of potential bearers of value, the primary examples of which are people and points in time. The challenge is to determine which category of value bearers are of ultimate moral significance: the \emph{ultimate locations}, for short. This paper defends the view that the ultimate locations are \emph{people at times}. A person at a time is not a specific person, but the person born at a  specific point in time (\emph{de dicto}). The main conclusion of the paper is that the unsettling implications of the time- and person-centered approaches to infinite ethics can be avoided.  
Most notably, a broad class of worlds that person-centered views deem incomparable can be strictly ranked. 
\end{abstract}
 
 \bigskip 
 \maketitle
 
\section{The locations problem}
\label{sect: intro}
 The locations problem  in infinite ethics concerns the moral status of people relative to points in time, states of nature, and other categories of potential bearers of value.  The challenge is to determine which category of value bearers are of ultimate moral significance: the \emph{ultimate locations}, for short.\footnote{Following \cite{Bro91}, the literature on infinite ethics   frequently refers to potential value bearers  as  \emph{locations}. The term \enquote{ultimate locations} was introduced by \cite{VK97} following the realization that  the prescriptions of  finitely additive infinite value theories typically depend on which locations are referenced.  I  use the term as  short hand for  \enquote{the locations referenced by the most plausible finitely additive infinite value  theory}.} I will deal with two aspects of this challenge. The first  %conOne aspect of this challenge  
concerns over which locations (i.e., bearers of value) the Pareto axiom is most appropriately defined. Given worlds  $w_1$ and   $w_2$ that have the same locations,    this axiom says that    $w_1$ is better than $w_2$   if some location has more utility in $w_1$  than in $w_2$, and no location has less utility in $w_1$.  It turns out that the prescriptions  of this axiom sometimes depend on which locations are referenced. We must then ask which locations \emph{should} be referenced.\footnote{This question presumes that we insist on \emph{some}  instantiation of Pareto, despite the concerns that have raised about doing so  \citep{van95,HM00b}.} 

The second aspect of the challenge concerns the notion of a locational order. In many of the proposed extensions of utilitarianism to infinite worlds, the  temporal (or spatio-temporal) order crucially determines the order  of summation. These pro\-posals include the PMU criterion of \cite{Val93},  Wilkinson's \citeyearpar{Wil21} spatio\-temporal version of expansionism, and almost all contributions to the economics literature stemming from the seminal work of \cite{Ram28}. While points in time arguably possess a natural order, there is near consensus that   % that there is no morally privileged way of ordering individuals.  
individuals cannot be  ordered in a natural way.  For utilitarians, the second aspect of the challenge might thus be cast as that of  deciding whether or not the relevant utilities should be summed in a particular order.
 
Some utilitarians see people as containers for well-being. On this view, what matt\-ers is not people, but rather the intensity and duration of subjective experiences.  Other authors, including Peter Vallen\-tyne, John Broome, Nick Bostrom and Amanda Askell, take the opposite viewpoint. As Vallen\-tyne writes, \enquote{the spirit of traditional utilitarianism goes with the person-centered  app\-roach rather than the time-centered approach}.\footnote{See \citet[417]{Val95}. \citet[19]{Bos11} and \citet[14]{Ask18} make similar remarks. \citet[Ch. 7]{Bro06} addresses a version of the locations problem in finite worlds. He rejects the time-based (or \enquote{snap-shot}) view in favor of the person-centered view.} Finally, some authors remain neutral as to whether the ultimate locations are people, points in time, or something else. For example, neither \cite{LV04} nor \cite{Arn14} take a stance on what the ultimate locations are.

It has long been known that, in infinite worlds, it makes a crucial difference which locations are referenced.\footnote{I will consider three types of locations: times, people, and what I  refer to as people at times. \S \ref{subsect: against times}  and \S \ref{subsect: against times2} illustrate the fact, due to \cite{Cai95}, that  Pareto over  times conflicts with  Pareto over people. We will also see that Pareto over people at times conflicts with both Pareto over times (\S \ref{subsect: against times2}) and Pareto over people (\S \ref{subsect: objection3}).}  As \cite{Cai95} demonstrates (and as I  illustrate below), Pareto over times conflicts with Pareto over persons.  If we insist that people are impor\-tant, as many people do, the time-centered approach thus seems untenable. 

The person-centered approach has a clear moral basis. However, as \cite{Ask18} shows and as we will later see, person-centeredness leads to generic incomparability:  \enquote{most} pairs of infinite worlds cannot be ranked. One might react to this result by concluding that infinite ethics is fundamentally undecidable. But we might also ask if it is possible to place significance on people in such a way that generic incomparability is not entailed.

In this paper, I defend the view that the ultimate locations are \emph{people at times}. Informally, a person at a time  is not a specific person, but the person born at some specific point in time (\emph{de dicto}). In \S \ref{sect: pat}, I give a formal definition of people at times using a counterpart relation that identifies individuals born at the same time in different futures. I show that, under certain assumptions (stated in \S \ref{sect: pat}), this counterpart relation can be deduced from Parfit's \citeyearpar{Par84}  No Difference View. In  \S \ref{sect: Ordeal and Depletion}--\ref{sect: comparability},  I explore the impli\-cations of taking people at times as the   locations.  My conclusion from this investigation is that the unsettling impli\-cations of the time- and person-centered approaches can be avoided. Most notably, a broad class of infinite worlds that   person-centered views deem incomparable can be strictly ranked.  
 
\section{Against times and people as ultimate locations} \label{sect: against times and people}
I begin by recalling  the primary criticisms of the time- and person-centered approaches to infinite ethics.  I return to these criticisms in \S \ref{sect: Ordeal and Depletion}. 

\subsection{Against times}\label{subsect: against times}
  The most influential critique of time-centeredness is due to \cite{Cai95}.  
 The following is a version of one of his thought experiments.\footnote{See \citet[401]{Cai95}.}

\begin{quotation}
\textbf{Ordeal}. At times $t = 1, 2, 3,\ldots$, a person is born and lives for two equally long periods. Each person's life is an ordeal where utility is negative in the first period and positive in the second period. As a whole, each person has a life well worth living. Table \ref{table: 1} displays the utility profile of an infinite world that obtains in this way. I will refer to this world as \emph{Ordeal}.
\end{quotation}
{\renewcommand{\arraystretch}{1.12}
\begin{table}[ht]
    \begin{center}
    \begin{tabular}{c|rrrrrrr|r}
          & $1$ & $2$ & $3$ & $4$ & $5$ & $\ldots$ & $\ldots$ &  sum \\\hline
       $p_1$  & $-2$ & $4$ & & & & & & $2$\\ 
       $p_2$ &  & $-8$ & $16$ & & & & &  $8$\\ 
       $p_3$    &  &   & $-32$ & $64$ & & & & $32$\\ 
       $p_4$  &   &  & & $-128$ & $256$ &  & & $128$\\ 
       $p_5$ &  &   &  &  & $-512$   & $\ddots$ & & $512$\\ $\vdots$ &  &   &  &  &  & $\ddots$ &  & $\vdots$ \\ \hline
      sum & $-2$ & $-4$  & $-16$ & $-64$ & $-256$    & $\ldots$ &  &   \\ 
      %\smallskip
    \end{tabular}
    \end{center}

    \caption{Row $i$ contains the utilities of person $p_i$ in two successive time periods of length one.   The  realized utility is negative in every period, yet the lifetime utility of every person is positive.    \label{table: 1} }
\end{table}}

In each time period, realized utility is negative.  So, summing over times would lead us to conclude that  Ordeal is   worse than the world where each person has zero   utility, at all times. But  intuition suggests that a world in which all lives are well worth living is better than a world in which everyone has utility $0$.  Many authors (myself included) have found this objection to time-centeredness to be decisive. 

\subsection{Against people}\label{subsect: against people}
One criticism of person-centered approaches says that such approaches are inadequate in  cases involving  \emph{future} people.  The following thought experiment from \citet[361]{Par84} is a well known illustration of this point. 
\begin{quotation}
\textbf{Depletion}. As a society, we must choose between conservation and depletion of a  natural resource. If we choose to deplete the resource, then the quality of life over the next few centuries will be slightly higher than what it would be if we had chosen to conserve the resource. However, if we choose depletion, then the quality of life will later, and then for many centuries, be much lower than what it would have been  had we chosen conservation.
\end{quotation}

As \citet[361]{Par84} argues,  we can plausibly assume that, after one or two centuries, there will be no one  in our society who would have been born whichever policy we choose. 
Depleting the resource is therefore not bad for any one specific person. In fact,  there is a person-centered argument for depleting the   resource: if we do, then some individuals  (those now living and those about to be born) will be better off, and  no-one will be worse off. Thus, at least some person-centered views seem to suggest that depletion is permissible.\footnote{\label{foot: person-affecting}To be sure, not all person-centered views deny that depletion is worse than conservation (cf. \cite{Mea12}).  However, depletion is permissible on any view that meets the person-affecting restriction which states  that a favored outcome  must be  favored by at least one individual.   This restriction has  received some heavy criticism,  but it is not without proponents---see and compare \cite{Hey09} and \cite{Tem93}  and \cite{Arr09}. For discussion, see  \cite{Rob19}. } This problem  (for these views) persists  in  the twist to Depletion that we obtain by replacing \enquote{for many centuries} above with \enquote{for eternity}.  We will see (in \S  \ref{subsect: Depletion}) that in the  case of an infinite time horizon, Depletion challenges a broader class of person-centered views. Indeed,  the person-centered principles that the literature on infinite ethics provides  cannot compare the outcome of choosing depletion with that of choosing conservation.

\section{People at times}\label{sect: pat}
The concept of  people at  times   will allow us to avoid the aforementioned criticisms of   time- and person-centeredness.  I will present this concept in a setting which is slightly more general than that of \cite{FM03}, but it will here be assumed that only one person is born at the start of each period.\footnote{\citet[799]{FM03} use a setting where there is a constant (finite) number of people within each one of a sequence of non-overlapping  time periods. In the special case when there is one person per period, their definition of a generation coincides with my informal definition of a person at a time.}   This assumption will make the key difference between person-centeredness and people-at-times-centeredness   easy to  flesh out, and % At the same time, it will 
it will still permit us to deal with many problem cases that have been  frequently discussed.  

Unless stated otherwise, I  assume the following: At time $t_1$, a single person comes into existence, lives for $l_1$ units of time and then dies. A new person comes into existence at time $t_2>t_1$, lives for $l_2$ units of time and then dies. And so on, without end.\footnote{Most of the discussion in this paper concerns the case when  $l_i=l_1$ and $t_i=i\cdot l_1$ for every $i$, but my arguments do not hang on these specifics.  In Ordeal (see \ref{subsect: against times}),     $t_i = i$ and $l _i= 2$.}   By a \emph{future}, I mean a sequence $w = ((u_1, p_1),  (u_2, p_2), (u_3, p_3), \ldots)$, where $u_{i}$ is the lifetime utility of $p_i$, the person born at time $t_i$.  I will  primarily be   interested in the  problem of  how utilitarianism  should rank infinite futures.  

In philosophy, this problem was first discovered  by \cite{Seg76}. It was rediscovered by \cite{Nel91}, and a solution was  proposed by \cite{Val93}. Beginning with Ramsey's  \citeyearpar{Ram28}  work on capital accumulation, the problem has also %attracted the attention  of  economists working in various areas 
been studied from the viewpoint of intertemporal  social choice, particularly in welfare economics and environmental economics.\footnote{See, e.g., \cite{Wei65},   \cite{vLL97}, \cite{FM03},  \cite{BM07},  \cite{Roe11}, and \cite{ZA12}.}   
 My primary aim   in this paper 
  is to contribute to the long-standing  debate in philosophy about which locations should be referenced. But the paper can also be interpreted as an attempt to provide a philosophical basis for the utility streams framework used in the literature on inter\-generational equity. %\footnote{An often used assumption in this literature is that there is a constant (finite) number of people within each one of a sequence of non-overlapping  time periods  \citep[see, e.g., ][799]{FM03}. In the special case when there is one person per period, the definition of a generation coincides with my informal definition of a person at a time.}  
  As \citet[12]{Ask18} points out, such a basis has not been provided even in the   case when one person lives at each point in time.  

Informally, a person at  a time is not a specific person, but the person born at some specific point in time (\emph{de dicto}). My first goal is to give a  formal definition of this concept. My definition uses a counterpart relation induced by Parfit's  \citeyearpar{Par84}   No Difference View (NDV).  Before I define people at times, I therefore need to recall what NDV says and explain its relevance   to the problem at hand.

% The No Difference View 
 NDV is Parfit's  response to forward-looking implications of the \emph{non-identity effect} in cases like Depletion.  The non-identity effect is the indirect or random effect that  acts  performed today have on the genetic makeup  (and therefore the identities) of future people. As already mentioned,   \citet[361]{Par84} argues that the selection of social or environmental  policies provide examples of identity-affecting acts.  It is widely believed  that a much wider class of acts eventually affect which sperm-egg combinations become living beings a few centuries from now. For instance,   \citet[350]{Gre16} argues that this class  %of identity-affecting acts  
 includes acts as mundane as   that  of helping someone across the street.  
  
 Due to the non-identity effect, we can thus  plausibly assume that  any two futures eventually have no individuals in common. 
For expositional brevity, I will assume  that any two futures have no individuals in common at all. %In particular, I assume that people cannot exist at radically different times in different futures. The      last assumption is  implied by Parfit's \citeyearpar[351]{Par84} Time Dependence Claim (TDC), which says that if a person had not been born within a month of when he or she was born, then he or she would never had existed.  
I   explore the implications of relaxing  this assumption in  \S \ref{subsect: objection3}. % by considering a thought experiment  where  TDC is false. 

 The No Difference View 
 tells us to reason as if our acts were \emph{not} identity-affecting. \citet[168]{Mul02b}  formulates NDV in the following way: 
\begin{quotation}
NDV (No Difference View). If we believe we are facing a same-people choice, and discover that it is actually a different-people choice, then this should make no difference to our moral deliberations.
\end{quotation}
This formulation of NDV could use a qualifier to clarify that the reason why we initially believed that we were facing a same-people case, but then \enquote{discovered} that we actually have a different-people case, is that  the choice itself is identity-affecting. With this qualifier,  NDV might be expressed as follows.
\begin{quotation}
If  a forward-looking 
different-people case would have been a same-people case had it not been for the non-identity effect, then the moral ranking is the same as in a same-people case.  
\end{quotation}
As   \citet[168]{Mul02b} notes, the No Difference View  is a feature of most utilitarian accounts of our obligations to future people.  As stated above, I am here interested in the problem of how utilitarianism should rank possible futures. %The No Difference View
NDV is therefore   relevant to the problem at hand. %in the context of the present discussion.  

My next step is to observe  that NDV induces a counterpart relation that identifies people born at the same time in different futures. To see this, consider two futures, $ ((u_1, p_1), (u_2, p_2), (u_3, p_3),$ $\ldots)$ and $ ((u_1', p_1'), (u_2', p_2'), (u_3', p_3'), \ldots)$. If there were no non-identity effect, then $p_t$ and $p_t'$ \emph{would be} the same individual. By NDV, we can therefore reason as if $p_t$ and $p_t'$ were the same individual.  Let $\rN$   denote the  counterpart relation that identifies $p_t$ and $p_t'$.\footnote{Note that the meaning of a counterpart in the present discussion is different compared with the  Counterpart Theory of David Lewis, where counterpart relations are similarity relations.  In the present discussion, that $p'$ is a counterpart to $p$ means that it is not morally relevant to distinguish $p$ and $p'$. This is not a claim to the effect that $p'$ is more similar to $p$ than to   other objects in the world that $p$ inhabits. I  return to this point  in \S \ref{subsect: objection1}.}

Note that if we identify people born at the same time in different futures, it becomes convenient to associate a future with a   sequence $(u_1, u_2, u_3,\ldots)$ of utilities. We then interpret $u_i$ not as the lifetime utility of a specific person, but 
as the  lifetime utility the person born at time $t_i$ \emph{de dicto}. This is one way of  interpreting the utility streams framework used in the  literature on intergenerational equity \citep{Dia65,FM03,Lau10}.% in the case when   one person lives at each point in time. 

I am now in a position to  define the concept of a person at a time in the setting described at the beginning of this section, 
where single people come into existence at arbitrary, fixed times. The binary relation $\rN$ is defined on the set of potential people. If  $ ((u_1, p_1), (u_2, p_2),$ $(u_3, p_3),$ $\ldots)$ and $((u_1', p_1'), (u_2', p_2'), (u_3', p_3'), \ldots)$ are possible futures, then $\rN$ identifies $p_t$ and $p_t'$,  for every $t$. That is,   $p \rN p'$  if    $p'$ and $p$ live at the same time in different futures.  This binary relation    is reflexive, symmetric and transitive (i.e., an equivalence relation). Its equivalence classes  consist of people born at the same time in different futures. These equivalence classes provide a formal definition of the concept of people at times. 

The rest of the paper explores the implications of letting people at times be the locations. During this investigation, I   use their informal definition.  Thus, if people at times  are referenced, then  
\begin{quotation}
\enquote{the person born at time $t$}
\end{quotation}
is interpreted \emph{de dicto}.  In this case, the locations are the same across alternatives. When people  are referenced, \enquote{the person born at time $t$} is interpreted \emph{de re}.   We will see that the subtle difference between people and people at times not only leads to conflicting versions of   Pareto.  It also bears on the  important question of whether   the relevant locations  come in a particular order.

\section{Ordeal and Depletion}\label{sect: Ordeal and Depletion}
In \S \ref{sect: against times and people}, I touched on some criticisms of the time- and person-centered approaches to infinite ethics. In this section, I  expand on these criticisms and show that the person-at-times-centered approach avoids them.  

A note on terminology: In the literature of infinite ethics,   a \enquote{person-centered} view is one where people are referenced as the locations. In the context of conflicting versions of Pareto, person-centered views give priority to Pareto over people. It should be noted that person-centeredness  does not imply any  commitment to \emph{person-affecting} restrictions.  In fact, person-centered views typically do not meet the restriction that a favored outcome must be favored by at least one individual.\footnote{Consider, for example, the  view that the sum of individual utilities should determine the ranking of all infinite worlds for which the relevant sums are well defined and absolutely convergent. This view is person-centered, but does not meet the person-affecting restriction   (see footnote \ref{foot: person-affecting}).}   With respect to person-affecting restrictions, there might thus appear to be little difference between person-centeredness  and person-at-times-centeredness. We will see though that Pareto over  people at times conflicts with Pareto over  people (see \S \ref{subsect: objection3}). So, although the view that I defend does place significance on people, it does not qualify as a person-centered view.   

\subsection{Ordeal: times vs people at times}\label{subsect: against times2}
Recall from \S \ref{subsect: against times} that  Ordeal is a world where   single people come into existence at times $t = 1, 2, 3, \ldots$ and live for two  time periods of   length one. 
The  utility realized in each period is negative, but the lifetime utility of each person is positive---see Table \ref{table: 1}. One version of the standard objection to time-centeredness points out that Pareto over times entails that Ordeal is worse than the world where everyone has utility $0$.

From the person-centered viewpoint, Ordeal is \emph{better} than the world where everyone has utility $0$. Indeed, in Ordeal, the person born at time $t$  \emph{de re} has a life well worth living, for  every  $t$.  It is also true that, for  every  $t$, the person born at time $t$  \emph{de dicto} has a life well worth living.  So the people-at-times-centered approach is not subject to the criticism that is often leveled  at time-centeredness.

Ordeal illustrates the difference between  time-centeredness and  people-at-times-centeredness. The difference between person-centeredness and people-at-times-centeredness is here more subtle. To illustrate, consider the choice between %bringing about 
 \begin{quotation}
$w$:  Ordeal, or 
 
 $w'$: a world where single people live at the same times as in Ordeal, but everyone has utility $0$, at all times.  
 \end{quotation} 
 If  $w$ and $w'$ contain the exact same individuals,\footnote{Some philosophers (notably \cite{Lew68}) deny that it is ever meaningful to speak of objects in different worlds as being the same.  On this view, a  counterpart relation can serve as a substitute for  transworld identity.  Readers who sympathize with this view may interpret my saying that  two individuals are the same to mean that they are counterparts in the least controversial sense possible. For an uncontroversial case, the reader is asked to think of a single-person world where the reader has coffee, and another  single-person world where the reader has tea.}  
 then $w$ is better  than $w'$ by Pareto over people. If, however, the choice \emph{is} identity-affecting, then Pareto over people is silent, because in this case the individuals  in $w$ and $w'$ are not the same. 
  In this case, $w$ is better than $w'$ by Pareto over people at times. Indeed, the person born at time $t$ \emph{de dicto} has more utility in $w$ than in $w'$, for every $t$.    
\subsection{Depletion: people vs people at times}\label{subsect: Depletion}
I now  return to % Eternal Depletion  (i.e., 
the twist to the Depletion case where depleting the resource increases the quality of life for a few centuries, but then  reduces it significantly for all eternity (see \S \ref{subsect: against people}). I will first show that people-at-times-centeredness leads to the conclusion that conservation is better than depletion. I then explain why person-centeredness leads to incomparability. 

It should be pointed out  that although incomparability is consistent with Askell's \citeyearpar{Ask18}  conclusion that most pairs of infinite worlds cannot be ranked,  the inability of person-centeredness to give a verdict in cases like Depletion is not widely recognized. 
\cite{Ask18} considers cases  where  locational order plays a crucial role. (I discuss such cases in the next section.) That person-centeredness  does not even allow us to deal with cases like Depletion is not well recognized in the literature on infinite ethics. 

Let us assume, for simplicity, that there is one person per period (say eighty  years), and that everyone in the conservation world, $w_C$, has utility $2$. In the depletion world, $w_D$, the  people   in the first two periods have utility $3$ and the rest have utility $1$. In other words,   the utility sequences that go with $w_C$ and $w_D$ are as follows:
\begin{align*}
w_C &:(2, \; 2, \; 2, \; 2, \, 2, \, 2,  \, 2, \ldots) \\
w_D &: (3, \; 3, \; 1, \; 1, \, 1,  \, 1,  \, 1,  \ldots).
\end{align*}
As in the original version of Depletion (where the time horizon is finite),      
 the two worlds eventually have no individuals in common.

With  people at times as the locations,  we can rank $w_C$ and $w_D$  using principles that enjoy relatively wide support. For instance,  
  the following principle entails that $w_C$ is better than $w_D$. 
 \begin{quotation}
SBI1. For worlds $w_1$ and $w_2$  that have the same locations,   if there are constants $c_1$ and $c_2$ with  $c_1>c_2$   such that    the utility of every location  in $w_1$ is greater than $c_1$ and the utility of the same location in $w_2$  is less than $c_2$,  except possibly for finitely many locations, then $w_1$ is better than $w_2$.	
\end{quotation}
This is Vallentyne and Kagan’s \citeyearpar[11]{VK97}  first strengthening of their Basic Idea.  Its person-centered instantiation is silent here, because the people in  $w_C$ and $w_D$ are not the same. By contrast, if people at times are the locations, then the locations are the same.  For every $t>2$, the utility of the person born at time $t$ \emph{de dicto}   is equal to $2$ in $w_C$, but   $1$ in $w_D$. So the people-at-times-centered version of % as the locations, %the  first strengthening of Vallentyne and Kagan’s \citeyearpar{VK97}  Basic Idea %the principle   
SBI1  says that  $w_C$ is better than $w_D$. 
% \newpage
 
As I mentioned  in \S \ref{sect: against times and people},    the person-centered principles that the literature provides  are  unable to  compare $w_C$ and $w_D$. This includes  the person-centered  principles of  \cite{VK97} and   \cite{LV04}\footnote{\label{foot: order infinite sum}Vallentyne and Kagan's \citeyearpar{VK97}   principles cover two types of  cases, one where locations are the same and one where locations have a natural order. Their first set of principles do not apply in   Depletion with people as the locations, because the people in $w_C$ and $w_D$ are not the same. Their second class of principles do not apply since  (they argue) people come in no particular order. Lauwers and Vallentyne's \citeyearpar{LV04} principles are silent in Depletion for the same reasons.} 
 and  %We will see that it 
 Arntzenius' \citeyearpar[55]{Arn14}  Weak People Criterion.\footnote{See footnote \ref{footnote: WPC} and the discussion preceding it.} 
The basic problem for person-centeredness in cases like Depletion is that, without sameness of locations, the mere  fact that the individual  (or location-wise) utility level is higher  in one   infinite world than in   another does not say much about if and how the  worlds should be  ranked.

For readers unfamiliar with the  pitfalls of infinite ethics, the last claim may  need explanation.  To clarify, consider two  worlds, $w_1$ and $w_2$,   populated by infinitely many individuals, not necessarily future people. The individuals in $w_1$  have utility $1$ while those in $w_2$ have utility $2$.  If $w_1$ and $w_2$ contain the same individuals, then $w_2$ is better than $w_1$ by Pareto over people.\footnote{If  $w_1$ and $w_2$ contained the same \emph{finite} number of people,   we would not need sameness or a counterpart relation to make plausible that $w_2$  is better than  $w_1$; in this case, Suppes dominance says that  $w_2$  is better   since there is a one-one correspondence between the populations of $w_1$ and   $w_2$ such that at least one person in $w_2$ is better off than the corresponding person in $w_1$, and every person in $w_2$ is at least as well off as the corresponding person in $w_1$ \citep[cf.][233]{Bro18}. We are about to see that,  unless the one-one correspondence in question is a plausible counterpart relation, this notion of dominance becomes implausible in infinite worlds.}  
However, in some cases it is implausible that $w_2$ is better than $w_1$. As \cite{Val95} notes, one problem is  that  $w_1$ may contain all locations from $w_2$, plus a hundred clones of each. On these grounds, he concludes that traditional utilitarianism should not, in general, judge  a  world with a higher  individual utility to be better than  a world with a lower individual utility.\footnote{\label{foot: Val95}See \citet[419]{Val95}.  The same reasoning leads \citet[20]{VK97} to  reject the idea that infinite worlds with a higher  individual utility  are in general better than  worlds with a lower individual utility.} For another reason to take this position, consider a future world where each generation consists of one billion people that each have utility $2$, and a second world where each generation consists of ten people, each with utility $2.1$. Then the individual utility level is higher in the second world. Yet from the   perspective of total utilitarianism, it seems highly doubtful that the second world is better. 

It is of course  \emph{possible} to define person-centered principles that \emph{do} rank $w_C$ above $w_D$. For example, anyone is free to declare one world to be better than another if the individual utility level is higher in  the first world, except possibly for finitely many individuals. But,  as we have  just  noted,  such a principle  becomes implausible  when we consider infinite worlds whose populations are not the same.  We might restrict it in such a way that it applies in Depletion, but is silent in other different-people cases. However, in doing so we will need to explain what distinguishes   Depletion  from other different-people cases. It seems that such explanations will ultimately have to appeal to the fact that Depletion involves  people whose existence is contingent on our acts.  We must then concede that  non-identity cases can be dealt with as   same-people cases,
 which is precisely what NDV asserts. And as we have seen, NDV  leads us away   from person-centeredness  towards the impersonal concept of people at times. 

Finally, let us note   that the inability of person-centered principles to deal with the Depletion case is not cured by identifying non-existence and existence with utility $0$. For suppose we do. We must then compare 
 \begin{quotation}
  world $w_C'$  where infinitely many individuals have utility $0$ (those with utility $1$ in $w_D$) and the infinitely many remaining   individuals  have utility $2$, and
 
   world  $w_D'$ where two  individuals  have utility $3$, infinitely many  have utility $0$, and the infinitely many remaining  individuals have utility $1$. 
  \end{quotation}
By assumption, these worlds contain  the same individuals. But since infinitely many of them have more utility in $w_C'$ while infinitely many have more utility in $w_D'$,  this does not help %aforementioned extension of Vallentyne and Kagan’s \citeyearpar{VK97} Basic Idea 
 (SBI1 is silent).  Neither does the fact that the sum of individual utility differences may approach plus infinity or minus infinity depending on how these differences are ordered.%Arntzenius' \citeyearpar[55]{Arn14}   criterion %Weak People Criterion (WPC) 
\footnote{\label{footnote: WPC}This is why Arntzenius' \citeyearpar[55]{Arn14}  Weak People Criterion (WPC) does not give a verdict. 
%\citet[160]{Ask18} formalizes WPC using the following notation: 
\citet[160]{Ask18} formalizes WPC %Arntzenius' \citeyearpar{Arn14} Weak People Criterion (WPC) 
using the following notation: 
Given worlds $w$ and $w'$, let $P$ be the set of  people that are in either $w$ or $w'$. If $p \in P$ exists in $w$, let $u_{w}(p)$ denote $p$'s utility in $w$; if not, then $u_{w}(p)=0$. WPC says that $w$ is better than $w'$ if and only if   $\sum_{i=1}^n(u_{w}(p_i)-u_{w'}(p_i))$ tends to a positive real number, or plus infinity, for every enumeration $p_1, p_2, p_3, \ldots$ of $P$.  To see that WPC does not rank $w_C$ above $w_D$,  note that  $\sum_{i=1}^n(u_{w_C}(p_i)-u_{w_D}(p_i))$  tends to minus infinity if we enumerate $P$ so that $p_i$ is in $w_C$ if and only if $i$ is, say, an integer multiple of $10$.} We \emph{could} get a verdict by placing significance on the individual's temporal positions.  However, as mentioned in the introduction,   doing so has generally been considered morally unjustified. I return to this aspect of the problem in the next section. 

\section{Incomparability vs  locational  order}\label{sect: comparability} 
So far, I have focused on  different versions of  Pareto.  Another much debated % equally significant %(and much debated) 
aspect of the problem at hand concerns the idea of a  locational order.  The  sticking point  is that many of the principles for ranking infinite worlds that the literature provides are sensitive to the order  in which utilities app\-ear.\footnote{These proposals include the \emph{medial limit} of \cite{vLL97}, the order-sensitive principles of \cite{VK97} and Bostrom's \citeyearpar{Bos11} \emph{value density}.} In this section, I will first describe the main objection to these principles. I then explain why the objection  is  not relevant when the locations are people at times. 

The following case has generated a considerable amount of attention.\footnote{The   case considered in this section first appears in \citet[235]{Seg76}. It has reappeared many times in different forms, but  not all authors are explicit about whether or not people are the same across alternatives.  Askell \citeyearpar{Ask18} explicitly assumes that people differ across alternatives. Indeed, the case considered here is identical to her  Mansion and Shack \citep[9]{Ask18}.   }  
\begin{quotation}
\textbf{Cycles}. In two future worlds,  $w_1$ and $w_2$,   single people live in non-overlapping time periods  (say 80 years). The living conditions in each period  are either   good, corresponding to utility $2$, or bad, corresponding to utility $1$.  In $w_1$, two good periods are followed by a bad period, which is followed by two good periods, and so on.  The periods that are good  in $w_1$ are bad in $w_2$, and vice versa. Thus, the utility profiles of $w_1$ and $w_2$ are as follows: 
\begin{align*}
    w_1 &:  (2, 2, 1, 2, 2, 1, 2, 2, 1,  \ldots)\\
   w_2 &: (1, 1, 2, 1, 1, 2, 1, 1, 2,     \ldots).
\end{align*}
As before, $w_1$ and $w_2$ have no individuals in common.
\end{quotation}

 Our first reaction is perhaps that $w_1$ is better than $w_2$ since $w_1$ has more utility   in   every  sufficiently large time interval.  As appealing as this intuition may seem, it presumes a locational order.  Once again, we   must   ask which locations should be referenced; in this case, what matters most is if they come in a  particular order.

Points in time have generally been considered to posses a natural order. So, with   times as the locations, we can compare $w_1$ and $w_2$ using principles that are sensitivity to the order in which utilities appear.  However, in the post-\cite{Cai95} era of infinite ethics, most authors deny that times are ultimate locations. There is a clear moral basis for taking people as the locations, but then there is no natural order.  As \citet[19]{Bos11} points out,  since people come in no particular order,  it is unclear if worlds like $w_1$ and $w_2$ can be strictly ranked.  \cite{Ask18}  further develops this position. She defends the view that people are the ultimate locations and argues that worlds like $w_1$ and $w_2$ are incomparable  (i.e.,  that they cannot  be strictly ranked, but  are not equally good).\footnote{For a lucid description of Askell's  \citeyearpar{Ask18}  argument, see \citet[1926-27]{Wil21}.}

Summing up, in order to avoid   incomparability in Cycles,   it  seems we must   appeal to a locational order.\footnote{\cite{Wil21}  avoids incomparability by taking 
points in time as the locations. %This allows him to invoke the   order-sensitive principles of  \cite{VK97} to produce strict rankings in many cases of interest. 
As he acknowledges, this proposal  remains susceptible to the standard criticism of time-centeredness  (see \S \ref{subsect: against times} and \ref{subsect: against times2}). Another way to avoid incomparability in Cycles would be to declare worlds like $w_1$ and $w_2$ equally good, a conclusion which has also been described as unsettling (see, e.g., \citet[235]{Seg76},   \citet[164]{vLL97}, and  \citet[1928]{Wil21}).  }  
%For instance, \citet[235]{Seg76} and \citet[164]{vLL97} describe  this conclusion as \enquote{absurd} and \enquote{quite unacceptable}, respectively, and  \citet[1928]{Wil21} makes similar remarks.  } 
The objection from person-centeredness says that we should place significance on people, and people come in no particular order.  This objection is difficult to deny, but it does not apply to people at times. As an illustration, consider the statement
\begin{quotation}
the person born at time $t$ could be born at some other time $t'$.  
\end{quotation}
This statement makes perfect sense if \enquote{the person born at time $t$} is interpreted \emph{de re}.\footnote{The  \emph{correctness} of the statement  depends,  among other things, on our conception of personal identity. Indeed, the statement is inconsistent with Parfit's \citeyearpar[351]{Par84} Time Dependence Claim. } On the  \emph{de dicto} reading,  the statement is internally inconsistent.  Likewise, although it is arguably true that people  come in no particular order, this is not so for people at times.\footnote{One might object that, while it is true that people at times come in a particular order, it is specific people  that matter, and people can appear in any order.  In \S \ref{subsect: objection3} we will  see that there is an air of contradiction about this argument. } 
Consequently, the objection to the   order-sensitive principles that the literature provides does not apply if the locations are people at times.  If these principles are applied with people at times as the locations,  then worlds like $w_1$ and $w_2$   can be strictly ranked.
 
\section{Objections}\label{sect: conclusion}
I have argued that  person-at-times-centeredness avoids the standard criticism of time-centeredness and  yields plausible prescriptions where   person-centeredness leads to incomparability. In this  section, I tie up some loose ends by answering three objections.

\subsection{First objection}\label{subsect: objection1}  
Inspired by the literature on intergenerational equity, I have used a framework where  individuals come into existence at fixed times. % \citep[cf.][799]{FM03}. 
The first objection is that %while this framework is flexible enough to deal with many problem cases from the literature,
we can easily find problem cases that  the  framework does not accommodate.  %hjelsede

%It is true that the framework   is artificially simple. To find a problem case that it does not handle, it is enough to imagine that the list of birth-times differ by one second between outcomes.  
Besides reasons of space, the main reason why I have  focused on  idealized cases is that this is often  an effective way to attack difficult  questions. A central question in infinite ethics concerns if and how generic  incomparability can be avoided. Person-centeredness leads to incompara\-bility in idealized cases, so it makes sense to deal with such cases first. We have  seen how incomparability can be avoided in the kind of idealized cases that have been most frequently discussed. Given this, it  would be unsurprising if incomparability can be avoided in other,  more complicated cases. It would, for example, be unsurprising if incomparability can be avoided in   Depletion  if people are born one second earlier in one outcome, or if two people are born at the start of each period (in both outcomes). Such cases call for a more fluid concept of people at times. But that will be a topic for  another time. %
%Besides reasons of space, my primary reason for  focusing on  idealized cases is that this is often  an effective way to attack difficult  questions. A central question in infinite ethics concerns if and how generic  incomparability can be avoided. Person-centeredness leads to incompara\-bility in idealized cases, so it makes sense to deal with such cases first. We have  seen how incomparability can be avoided in the kind of idealized cases that have been most frequently discussed. Given this, it  would be unsurprising if incomparability can be avoided in  more complicated cases. It would, for example, be unsurprising if incomparability can be avoided in   Depletion  if people are born one second earlier in one outcome, or if two people live in each period (in both outcomes). Such cases call for a more fluid concept of people at times. But that will be a topic for  another time. %

\subsection{Second objection}\label{subsect: objection2}  As  already noted (in \S \ref{sect: pat}), the role that counterpart relations play in the  present theory is different compared with   David Lewis'  Counterpart Theory (CT), where counterpart relations are similarity relations. Lewis uses counterpart relations to clarify the meaning of  \emph{de re}  modal statements about specific objects. For example, in CT,   to claim that \enquote{I could have been a historian}  is to claim that there is a possible world where I have a counterpart who is a historian. By using counterpart relations in this way, Lewis  avoids having to require an object to both have and   not have particular properties.

The second objection says that counterpart relations  should be similarity relations.  This  is a legitimate  objection to any counterpart-based account of \emph{de re} modality. (That I have a counterpart who is an historian is no reason to proclaim that \enquote{I could have been an historian} if my counterpart is  not suitably similar to me.)  But the present theory has a different scope.  Here the counterpart relation serves to ensure that it is more relevant to compare $p$ with $q$  than it is to compare $p$ with   some other object in the world that $q$ inhabits.

This response does not explain why we should identify individuals born at the same time in different futures, as the equivalence relation   $\rN$ requires. A formal derivation of this  relation  was provided in \S \ref{sect: pat}  under the assumptions stated there.  It can at least in part be understood intuitively by noting that the only way in which we really can affect the well-being of future people is through their living conditions. Living conditions change with time and we can sometimes affect how they evolve, as in Depletion.  This makes it  natural to think of \enquote{the people living in year $t$}   impersonally  in moral reasoning, as we do in everyday life.

The last claim needs substantiation. To elaborate, consider living conditions of two kinds,   $g$ and $b$, where $g$ gives  present people higher well-being.  As \citet[230-233]{Bro18} notes, in the case of future people  we must make assumptions about their preferences.  We  will never know if they   prefer coffee to tea. But when it comes to things like the amount of available  resources, the state of the environment or  the prevalence of disease, we can expect their  preferences to be similar to ours. Let  $g$ and $b$ be such that we can confidently assume that future people will  prefer $g$ to $b$. We are asked to compare two futures, $w_g$ and $w_b$,  which differ only in that during some time period $t$, the conditions in $w_g$ are of type $g$ and the conditions in $w_b$ are of type $b$. We could then quite plausibly hold that $w_g$ is better than $w_b$. We would not say that this is so  because  $g$ is  \emph{intrinsically} better.   We would say that $w_g$ is better since $g$ is better for the person that lives in  period $t$ (\emph{de dicto}).  
  
 \subsection{Third objection} 
\label{subsect: objection3}
The third objection points out that although  people-at-times-centeredness  avoids the standard criticism of time-centeredness (see \S \ref{subsect: against times2}), it may still conflict with Pareto over people.  I will reply  by pointing to a questionable implication of Pareto over people in cases where the conflict arises. More precisely, I will show that if we insist on Pareto over people, then the order in which people live becomes significant. Recall % (\S \ref{sect: comparability}) 
that one argument from person-centeredness insists that order-dependence  be avoided (see \S \ref{sect: comparability}). We will thus see that this argument   carries an air  of self-defeat.  

Before going into details, let me mention why a rejection of the person-centrered Pareto axiom  should perhaps not come as a surprise. In the case of present people, the idea that an outcome must be favored if it is preferred by some and dispreferred by no-one  has   strong intuitive appeal. In same-people cases, this idea  is   consis\-tent with total utilitarianism. But the same can be said of the person-affecting restriction which states (in slogan form) that a favored outcome must be favored by someone. This restriction has received criticism in non-identity cases that many have taken to be decisive.\footnote{See footnote \ref{foot: person-affecting}.}   In a similar way, we might find reasons to reject the person-centered version of Pareto  when we consider future people.

Since Pareto over people is silent unless  people are the same in the worlds that we compare, we need a case where this is so for the conflict to arise.  Let us therefore consider   the following thought experiment from \citet[13]{Ask18}.
\begin{quotation}
 \textbf{Freezer} A freezer contains infinitely many fertilized eggs, $e_1, e_2, e_3, \ldots,$ which can be incubated at any point in time. The person that $e_i$ becomes, once  incubated,  lives for   eighty years on an island that has room for one person at a time. The  living conditions  are either good, corresponding to utility $2$, or poor, corresponding to utility $1$.
\end{quotation}

Let $p_i$  be the person that $e_i$ becomes, once incubated. If we assume (as \citeauthor{Ask18} does) that the time of  incubation does not affect  $p_i$'s  identity, then   we can create different futures containing the exact same people.\footnote{One might object that the person that $e_i$ becomes \emph{will} in fact depend on the time of incubation, and that we therefore cannot create different futures with the exact same people.  Since I have assumed that different futures have no individuals in common, the objection   only  makes  this assumption less restrictive.  }  Thus,  in this thought experiment, the assumptions from \S \ref{sect: pat} do not hold. In particular,   in this thought experiment, %since a person can exist at radically different times in different futures, 
the Time Dependence Claim \citep[351]{Par84}   is false.
 
To see that Pareto over people here conflicts  with Pareto over people  at times, % in this setting,  
suppose  we can place the island in one  of two states, $A$ and $B$. In   $A$, %  living conditions are good   in even-numbered periods and  poor in odd-numbered periods;
 the  living conditions in period $t$ are good  only if $t$ is an integer multiple of $2$; in   $B$, the   conditions in period $t$ are good  only if $t$ is an integer multiple of $4$---see Table \ref{table: 2a}. 
 \begin{table}[h]
    \begin{center}
    \begin{tabular}{lrrrrrrrrr}
       period  & 1 & 2 & 3 & 4 & 5 & 6 & 7 & 8 & \ldots \\ \hline
       $A$-utility: & $1$ & $2$ & $1$ & $2$ & $1$ & $2$ & $1$ &  $2$ & \ldots \\ 
       $B$-utility:  & $1$ & $1$ & $1$ & $2$ & $1$ & $1$ & $1$ &  $2$ & \ldots \\ 
      \smallskip
    \end{tabular}
      \end{center}
        \caption{Time-series of utilities in state $A$ and state $B$.     \label{table: 2a} }
\end{table} 

\noindent 
Let $w$ be the world that obtains if   the island is in state $A$ and   $e_{t}$  is incubated in period $t, \, t=1, 2, 3, \ldots$, and let $w^\ast$ be the world that    obtains if  the island is in state $B$ and  the embryos are incubated in some other order.  Since   each state has infinitely many good periods  and infinitely many bad periods, the alternative order can be chosen so  that $w^\ast$ is better than $w$ by Pareto over people. By Pareto over people at times, on the other hand, $w^\ast$ is  worse than $w$ (see  Table \ref{table: 2a}). %, because the person   in period $t$ \emph{de dicto}  has more utility in $w$, %  (if $t$ is even, but not an integer multiple of $4$),  
%or the same utility in $w$ and $w^\ast$. % (see  Table \ref{table: 2a}).  
This is the third objection to  my proposal.  

My counter-objection is that one might   plausibly hold that it should not matter in which order  the embryos are defrosted. Pareto over people has the peculiar implication that the order   \emph{does} matter, even if   living conditions   remain unchanged. To see this, fix the island in state $A$ and let $w$ be the world that  we get by defrosting $e_t$ at the start of period $t, t = 1, 2, 3, \ldots$,  precisely as above. By  defrosting the embryos in a different order,  we can create a world, call it $w^{\ast\ast}$, such that one person (say $p_1$) is better off in $w^{\ast\ast}$ than in $w$, and no-one is worse off---see Table \ref{table: 2b}. Then $w^{\ast\ast}$ is better than $w$ by   Pareto  over people.
\begin{table}[h]
    \begin{center}
    \begin{tabular}{lrrrrrrrrr}
       period  & 1 & 2 & 3 & 4 & 5 & 6 & 7 & 8 & \ldots \\  \hline
      utility  &  $1$ & $2$ & $1$ & $2$ & $1$ & $2$ & $1$ & $2$ & \ldots \\ 
       $w$: & $e_1$ & $e_2$ & $e_3$ & $e_4$ & $e_5$ & $e_6$ & $e_7$ &  $e_8$ & \ldots \\ 
       $w^{\ast\ast}$:  & $e_3$ & $e_1$ & $e_5$ & $e_2$ & $e_7$ & $e_4$ & $e_9$ &  $e_6$ & \ldots \\ 
      \smallskip
    \end{tabular}
      \end{center}
        \caption{Two worlds obtained by incubating the embryos in two different orders.   \label{table: 2b} }
\end{table}

From the people-at-times-centered viewpoint, the order in which the embryos are defro\-sted does not matter, because the order will not affect the utility  of the person born at time $t$ \emph{de dicto}. So we again have a conflict between   person-centeredness   and  people-at-times-centeredness.  From the person-centered viewpoint, both conflicts can be seen as troubling for the concept of people at times. But the conflicts can also be interpreted as  more trouble for person-centeredness. On this interpretation, what matters is that the world is populated by people  that experience high well-being; who these people are and the order in which they appear is insignificant.

  \section{Concluding remarks} 
The question of how utilitarianism should deal with  an infinite future goes back to \cite{Ram28}. 
     \cite{Cai95} showed that we  get diffe\-rent answers depending on which  value bearers are given ultimate moral significance. Two conflicting views subsequently emerged: the time-centered  view and the person-centered view. From  the time-centered viewpoint, utilities are naturally ordered according to the times when they are realized. The person-centered view denies that there is a natural order. In doing so, it severely limits our discriminatory power.  In fact, on contempo\-rary accounts of person-centeredness, most pairs of infinite worlds are incomparable. I have argued that   this conclusion can be avoided if we place significance on people in an impersonal way.  I have developed this argument in the non-overlapping generations model of \cite{FM03}.  The argument  might thus help guide our moral  reasoning on a variety of economic, social and environmental issues in the context of intergenerational equity.

\bibliographystyle{chicago}
\bibliography{addition}

\newcommand{\noop}[1]{}
\begin{thebibliography}{}

\bibitem[\protect\citeauthoryear{Arntzenius}{Arntzenius}{2014}]{Arn14}
Arntzenius, F. (2014).
\newblock Utilitarianism, decision theory and eternity.
\newblock {\em Philosophical Perspectives\/}~{\em 28\/}(1), 31--58.

\bibitem[\protect\citeauthoryear{Arrhenius}{Arrhenius}{2009}]{Arr09}
Arrhenius, G. (2009).
\newblock Harming future persons.
\newblock In M.~Roberts and D.~Wasserman (Eds.), {\em Harming Future Persons},
  pp.\  289--314. Ashgate.

\bibitem[\protect\citeauthoryear{Askell}{Askell}{2018}]{Ask18}
Askell, A. (2018).
\newblock {\em Pareto Principles in Infinite Ethics}.
\newblock Ph.\ D. thesis, New York University.

\bibitem[\protect\citeauthoryear{Basu and Mitra}{Basu and Mitra}{2007}]{BM07}
Basu, K. and T.~Mitra (2007).
\newblock Utilitarianism for infinite utility streams: A new welfare criterion
  and its axiomatic characterization.
\newblock {\em Journal of Economic Theory\/}~{\em 133\/}(1), 350--373.

\bibitem[\protect\citeauthoryear{Bostrom}{Bostrom}{2011}]{Bos11}
Bostrom, N. (2011).
\newblock Infinite ethics.
\newblock {\em Analysis and Metaphysics\/}~{\em 10}, 9--59.

\bibitem[\protect\citeauthoryear{Broome}{Broome}{1991}]{Bro91}
Broome, J. (1991).
\newblock {\em Weighing Goods: Equality, Uncertainty and Time}.
\newblock Wiley-Blackwell.

\bibitem[\protect\citeauthoryear{Broome}{Broome}{2006}]{Bro06}
Broome, J. (2006).
\newblock {\em Weighing Lives}.
\newblock Clarendon Press.

\bibitem[\protect\citeauthoryear{Broome}{Broome}{2018}]{Bro18}
Broome, J. (2018).
\newblock Efficiency and future generations.
\newblock {\em Economics and Philosophy\/}~{\em 34}, 221--241.

\bibitem[\protect\citeauthoryear{Cain}{Cain}{1995}]{Cai95}
Cain, J. (1995).
\newblock Infinite utility.
\newblock {\em Australasian Journal of Philosophy\/}~{\em 73\/}(5), 401--404.

\bibitem[\protect\citeauthoryear{Diamond}{Diamond}{1965}]{Dia65}
Diamond, P.~A. (1965).
\newblock The evaluation of infinite utility streams.
\newblock {\em Econometrica\/}~{\em 33\/}(1), 170--177.

\bibitem[\protect\citeauthoryear{Fleurbaey and Michel}{Fleurbaey and
  Michel}{2003}]{FM03}
Fleurbaey, M. and P.~Michel (2003).
\newblock Intertemporal equity and the extension of the {R}amsey criterion.
\newblock {\em Journal of Mathematical Economics\/}~{\em 39\/}(7), 777--802.

\bibitem[\protect\citeauthoryear{Greaves}{Greaves}{2016}]{Gre16}
Greaves, H. (2016).
\newblock {C}luelessness.
\newblock {\em Proceedings of the {Aristotelian} Society\/}~{\em 116\/}(3),
  311--339.

\bibitem[\protect\citeauthoryear{Hamkins and Montero}{Hamkins and
  Montero}{2000}]{HM00b}
Hamkins, J.~D. and B.~Montero (2000).
\newblock With infinite utility, more needn't be better.
\newblock {\em Australasian Journal of Philosophy\/}~{\em 78\/}(2), 231--240.

\bibitem[\protect\citeauthoryear{Heyd}{Heyd}{2009}]{Hey09}
Heyd, D. (2009).
\newblock The intractability of the nonidentity problem.
\newblock In M.~Roberts and D.~Wasserman (Eds.), {\em Harming Future Persons},
  pp.\  3--25. Ashgate.

\bibitem[\protect\citeauthoryear{Lauwers}{Lauwers}{2010}]{Lau10}
Lauwers, L. (2010).
\newblock Ordering infinite utility streams comes at the cost of a non-{R}amsey
  set.
\newblock {\em Journal of Mathematical Economics\/}~{\em 46\/}(1), 32--37.

\bibitem[\protect\citeauthoryear{Lauwers and Vallentyne}{Lauwers and
  Vallentyne}{2004}]{LV04}
Lauwers, L. and P.~Vallentyne (2004).
\newblock Infinite utility: more is always better.
\newblock {\em Economics and Philosophy\/}~{\em 20\/}(2), 307--330.

\bibitem[\protect\citeauthoryear{Lewis}{Lewis}{1968}]{Lew68}
Lewis, D.~K. (1968).
\newblock Counterpart theory and quantified modal logic.
\newblock {\em Journal of philsoophy\/}~{\em 65\/}(5), 113--126.

\bibitem[\protect\citeauthoryear{Meacham}{Meacham}{2012}]{Mea12}
Meacham, C. J.~G. (2012).
\newblock Person-affecting views and saturating counterpart relations.
\newblock {\em Philosophical Studies\/}~{\em 158}, 257--287.

\bibitem[\protect\citeauthoryear{Mulgan}{Mulgan}{2002}]{Mul02b}
Mulgan, T. (2002).
\newblock Transcending the infinite utility debate.
\newblock {\em Australasian Journal of Philosophy\/}~{\em 80\/}(2), 164--177.

\bibitem[\protect\citeauthoryear{Nelson}{Nelson}{1991}]{Nel91}
Nelson, M. (1991).
\newblock Utilitarian eschatology.
\newblock {\em American Philosophical Quarterly\/}~{\em 28\/}(4), 339--347.

\bibitem[\protect\citeauthoryear{Parfit}{Parfit}{1984}]{Par84}
Parfit, D. (1984).
\newblock {\em Reasons and Persons}.
\newblock Oxford: Oxford University Press.

\bibitem[\protect\citeauthoryear{Ramsey}{Ramsey}{1928}]{Ram28}
Ramsey, F.~P. (1928).
\newblock A mathematical theory of saving.
\newblock {\em The Economic Journal\/}~{\em 38}, 543--559.

\bibitem[\protect\citeauthoryear{Roberts}{Roberts}{2019}]{Rob19}
Roberts, M.~A. (2019).
\newblock The nonidentity problem.
\newblock In E.~N. Zalta (Ed.), {\em The Stanford Encyclopedia of Philosophy}.
  Metaphysics Research Lab, Stanford University.

\bibitem[\protect\citeauthoryear{Roemer}{Roemer}{2011}]{Roe11}
Roemer, J. (2011).
\newblock The ethics of distribution in a warming planet.
\newblock {\em Environmental Resource Economics\/}~{\em 48\/}(3), 363--390.

\bibitem[\protect\citeauthoryear{Segerberg}{Segerberg}{1976}]{Seg76}
Segerberg, K. (1976).
\newblock A neglected family of aggregation problems in ethics.
\newblock {\em Nous\/}~{\em 10\/}(3), 221--247.

\bibitem[\protect\citeauthoryear{Temkin}{Temkin}{1993}]{Tem93}
Temkin, L. (1993).
\newblock Harmful goods, harmless bads.
\newblock In R.~G. Frey and C.~W. Morris (Eds.), {\em Value, welfare and
  morality}, Volume~3, Chapter~20, pp.\  291--324. Cambridge University Press.

\bibitem[\protect\citeauthoryear{Vallentyne}{Vallentyne}{1993}]{Val93}
Vallentyne, P. (1993).
\newblock Utilitarianism and infinite utility.
\newblock {\em Australasian Journal of Philosophy\/}~{\em 71}, 212--217.

\bibitem[\protect\citeauthoryear{Vallentyne}{Vallentyne}{1995}]{Val95}
Vallentyne, P. (1995).
\newblock Infinite utility: Anonymity and person-centeredness.
\newblock {\em Australasian Journal of Philosophy\/}~{\em 73}, 413--420.

\bibitem[\protect\citeauthoryear{Vallentyne and Kagan}{Vallentyne and
  Kagan}{1997}]{VK97}
Vallentyne, P. and S.~Kagan (1997).
\newblock Infinite value and finitely additive value theory.
\newblock {\em The Journal of Philosophy\/}~{\em 94\/}(1), 5--27.

\bibitem[\protect\citeauthoryear{Van~Liedekerke}{Van~Liedekerke}{1995}]{van95}
Van~Liedekerke, L. (1995).
\newblock Should utilitarians be cautious about an infinite future?
\newblock {\em Australasian Journal of Philosophy\/}~{\em 73\/}(3), 405--407.

\bibitem[\protect\citeauthoryear{Van~Liedekerke and Lauwers}{Van~Liedekerke and
  Lauwers}{1997}]{vLL97}
Van~Liedekerke, L. and L.~Lauwers (1997).
\newblock Sacrificing the patrol: utilitarianism, future generations and
  infinity.
\newblock {\em Economics and Philosophy\/}~{\em 13\/}(2), 159--174.

\bibitem[\protect\citeauthoryear{von Weizs\"{a}cker}{von
  Weizs\"{a}cker}{1965}]{Wei65}
von Weizs\"{a}cker, C.~C. (1965).
\newblock Existence of optimal programs of accumulation for an infinite time
  horizon.
\newblock {\em Review of Economic Studies\/}~{\em 32}, 85--104.

\bibitem[\protect\citeauthoryear{Wilkinson}{Wilkinson}{2021}]{Wil21}
Wilkinson, H. (2021).
\newblock Infinite aggregation: expanded addition.
\newblock {\em Philosophical Studies\/}~{\em 178\/}(6), 1917--1949.

\bibitem[\protect\citeauthoryear{Zuber and Asheim}{Zuber and
  Asheim}{2012}]{ZA12}
Zuber, S. and G.~B. Asheim (2012).
\newblock Justifying social discounting: The rank- discounted utilitarian
  approach.
\newblock {\em Journal of Economic Theory\/}~{\em 147\/}(4), 1572--1601.

\end{thebibliography}

\end{document}